\documentclass[amsmath,amssymb,prb,letterpaper,
  twocolumn, floatfix, superscriptaddress,longbibliography]{revtex4-2}
\usepackage{color}
\usepackage{graphicx}
\usepackage{dcolumn}
\usepackage{bm}
\usepackage{braket}

\usepackage[colorlinks=true, linkcolor=blue, urlcolor=blue,
            citecolor=blue]{hyperref}

\hypersetup{
 colorlinks,linkcolor=blue,citecolor=blue,urlcolor=blue}

\begin{document}

\preprint{APS/123-QED}

\title{Efficient Chebyshev polynomial approach to quantum conductance calculations: Application to twisted bilayer graphene}
 
\author{Santiago Giménez de Castro}
\affiliation {School of Physics, Engineering and Technology and York Centre for Quantum Technologies, University of York, York YO10 5DD, United Kingdom}
\affiliation {School of Engineering, Mackenzie Presbyterian University, S\~ao Paulo - 01302-907, Brazil}
\affiliation {MackGraphe – Graphene and Nanomaterials Research Institute, Mackenzie Presbyterian University, S\~ao
Paulo -01302-907, Brazil}
\author{Aires Ferreira}
\email{aires.ferreira@york.ac.uk}
\affiliation {School of Physics, Engineering and Technology and York Centre for Quantum Technologies, University of York, York YO10 5DD, United Kingdom}
\author{D. A. Bahamon}%
\email{dario.bahamon@mackenzie.br}
\affiliation {School of Engineering, Mackenzie Presbyterian University, S\~ao Paulo - 01302-907, Brazil}
\affiliation {MackGraphe – Graphene and Nanomaterials Research Institute, Mackenzie Presbyterian University, S\~ao
Paulo -01302-907, Brazil}

\begin{abstract}
In recent years, Chebyshev polynomial expansions of tight-binding Green's functions have been successfully applied to the study of a wide range of spectral and transport properties of materials. However, the application of the Chebyshev approach to the study of quantum transport properties of noninteracting  mesoscopic systems with leads has been hampered by the lack of a suitable Chebyshev expansion of Landaeur's formula or one of its equivalent formulations in terms of Green's functions in Keldysh's perturbation theory. Here, we tackle this issue by means of a hybrid approach that combines the efficiency of Chebyshev expansions with the convenience of complex absorbing potentials to calculate the conductance of two-terminal devices in a computationally expedient and accurate fashion. The versatility of the approach is demonstrated for mesoscopic twisted bilayer graphene (TBG) devices with up to $2.3\times10^6$ atomic sites. Our results highlight the importance of moiré effects, interlayer scattering events and twist-angle disorder in determining the conductance curves in devices with a small twist angle near the TBG magic angle $\theta_m \approx 1.1^\circ$.

\end{abstract}

\maketitle

\section{\label{sec:Introduction}Introduction}

The quantum scattering model due to Landauer \cite{Landauer_1989} has become a central tool in mesoscopic physics because it allows for a clear interpretation of phase-coherent electron transport in terms of a transmission problem \cite{Imry_Landaeur_RevModPhys.71.S306}. Within this framework, the conductance of a mesoscopic system coupled to ideal leads reads $G=(e^2/h) \sum_{n,m} \mathcal{T}_{nm}$, where  $\mathcal{T}_{nm}$ is the transmission probability to scatter elastically across the system from channel $n$ on the source lead to channel $m$ on the drain lead. Thus, for an ideal conductor
, the Landauer formula predicts that changes in low-temperature conductance  occurs in discrete steps of $e^2/h$ (per spin) each time a new transport channel becomes accessible at the Fermi level. This unique fingerprint of noninteracting one-dimensional (1D) conductors was first observed in semiconductor ballistic point contacts more than thirty years ago \cite{Exp_van_Wees_88,Exp_Wharam_1988}, and subsequently in a variety of systems, including nanowires \cite{Exp_Lu_05,Exp_Weperen_13,Exp_Gooth_17}, carbon nanotubes \cite{Exp_Frank_98,Exp_Liang_01,Exp_Javey_04,Exp_Biercuk_05} and graphene devices \cite{Exp_Tombros_2011,Exp_Terres_16}.

Meanwhile, the development of efficient tight-binding frameworks for numerical  quantum transport simulation has been receiving considerable interest because they can be used to handle realistic geometries as well as to elucidate the role of imperfections and disorder  \cite{Szafer_Stone_89, Marel_89,Kirczenow_89,Takagaki_92,Maslov_Stone_95,Kirczenow_18}. Among these, tight-binding Green's function (TBGF) methods have become a standard class of tools owing to their flexibility and computational efficiency \cite{KWANT_2014,TBTK_2019,KITE_2020,FAN20211}. In addition to providing a convenient framework to calculate the conductance in multi-terminal devices, the TBGF approach allows determination of current distributions, local density of states and other quantities of interest, and can be extended to incorporate the effect of interactions \cite{datta1997electronic,ferry1999transport}. Notwithstanding its proven merits, the standard implementations of the TBGF method suffer from cubic algorithmic complexity, which severely limits the system sizes attainable. The popular recursive Green's function (RGF) technique \cite{Lee_81,Lewenkopf_13,PhysRevB.77.115119} partly mitigates this issue by partitioning the computational domain into small unit transverse sections whose Green's functions are recursively generated, but still requires the inversion of matrices whose size scale with the width of the transverse section. This technical challenge has not precluded the study of ballistic transport through a variety of nanostructures (including quantum dots \cite{Rotter_00}, interfaces between bulk crystals \cite{Wortmann_02}, and disordered topological insulators \cite{Focassio_20}), but presents a significant hurdle for performing large-scale simulations beyond the ballistic regime as well as for tackling complex devices composed of many different materials or with sub-units displaying large unit cells.  

In this paper, we revisit the linear-response transport framework and formulate a Chebyshev polynomial-based spectral technique that will allow us to bypass altogether expensive matrix inversions in the numerical evaluation of the conductance of mesoscopic systems. The approach, which makes use of a complex absorbing potential (CAP) to alleviate the computational resources needs \cite{Muga/2004, Manolopoulos/2004}, is applied to two-terminal \textit{twisted bilayer graphene} (TBG) devices containing in excess of a million orbitals. Our results show that the spatial modulation of the interlayer couplings that is responsible for the dramatic modification of the band structure of TBG \cite{TBLG_tw_Carr2017,TBLG_tw_Cao2018,TBLG_tw_Isobe2018,TBLG_tw_Ribeiro2018,TBLG_tw_Andrei_20} translates into important features in the conductance curves, including the appearance of symmetric peaks located at energies of the van Hove singularities which merge into a single peak (centered at zero energy) as the twist angle approaches a magic angle. This article is structured as follows: Section \ref{sec:Methodology} lays out the spectral approach to calculating the two-probe conductance. We also discuss the CAP strategy employed to handle the leads efficiently and show how it can be implemented by means of a simple modification of the Chebyshev recursion relations. Section \ref{sec:TBG} presents our results for the ballistic transport regime of TBG nanoribbons with a $ 3\times 10^4$ nm$^2$ cross-section area, which is, to our knowledge, the largest such system simulated with a real-space TBGF method so far. Section \ref{sec:Disorder} investigates the impact of twist-angle disorder in the quantum transport properties. Our results are summarized in Sec. \ref{sec:Conclusions}.

\section{\label{sec:Methodology}Model and methods}

\begin{figure}
\begin{center}
\resizebox{8.6cm}{!} {\includegraphics{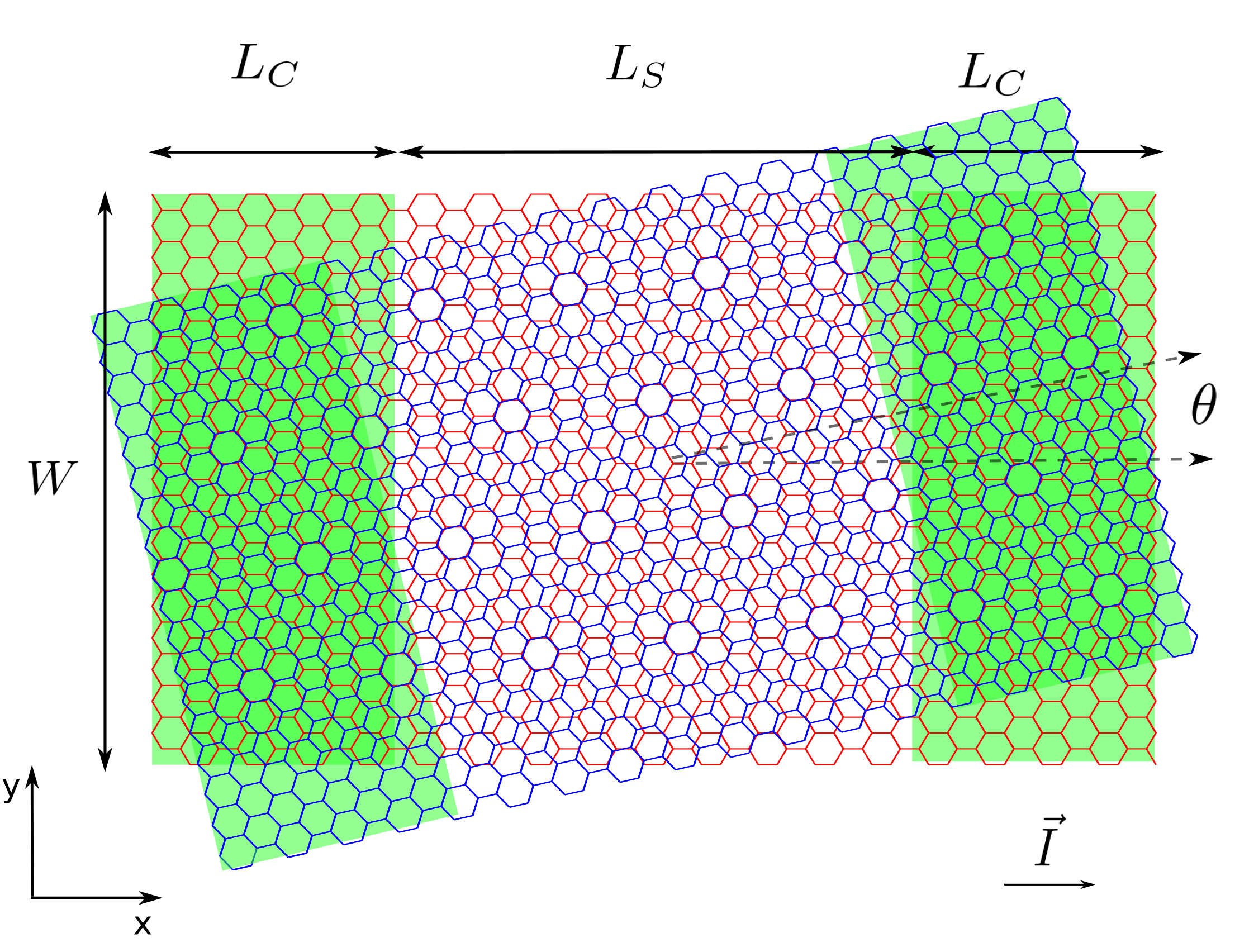}}
\caption{Two-terminal TBG device considered in this work. Green regions denote the leads. The largest systems considered here have $W=100$ nm and $L=L_S+2L_C=300$ nm, corresponding to a total of $~2.2$ million orbitals.}
\label{fig:schemeTBG}  
      \end{center}  
 \end{figure}

We consider a two-terminal device setup composed of a central region of length $L_S$ connected to leads of length $L_\text{C}$ (Fig. \ref{fig:schemeTBG}). The corresponding atomistic tight-binding Hamiltonian may be written as 
\begin{equation}
\widehat{H}=\begin{pmatrix} \widehat{H}_L & \widehat{V}_L &  0 \\ \widehat{V}^{\dagger}_L & \widehat{H}_C & \widehat{V}_R \\ 0 & \widehat{V}^{\dagger}_R & \widehat{H}_R \end{pmatrix},
\label{eq:Hamiltonian}
\end{equation}
\noindent where $\widehat{H}_{R(L)}$ and $\widehat{H}_C$ are the Hamiltonians of the right (left) lead and central region, respectively, and $\widehat{V}_{R(L)}$ describes the coupling of right (left) contacts to the central region. 
 
The linear-response conductance is obtained via the Kubo-Greenwood formula
\begin{equation}
  G(E) = \frac{4e^2}{h L_S^2} \textrm{Tr}\left[\hbar \hat{v}_x\hspace{1mm} \textrm{Im}\,\widehat{\mathcal G}(E)  \hspace{1mm}  \hbar \hat{v}_x  \hspace{1mm} \textrm{Im}\, \widehat{\mathcal G} (E) \right], 
\label{eq:KuboEq}
\end{equation}
\noindent where $E$ is the Fermi energy, $\hat{v}_x=(i/\hbar)[\widehat{H}_C,\hat{x}]$  is the velocity operator in the \textit{x} direction and  $\widehat{\mathcal G}=(E-\widehat{H}+i0^+)^{-1}$ is the TBGF of the full system. Note that $\hat{v}_x$ has support only on sites within the central region, so that Eq.~(\ref{eq:KuboEq}) correctly describes the total electric current ($I$) flowing in response to constant voltages applied at its boundaries.  The linear-response formulation is preferred here over the more commonly employed non-equilibrium Keldysh technique \cite{Caroli_1971,Meir_1992} since it is amenable to a spectral representation in terms of Chebyshev polynomials similar to the bulk longitudinal conductivity \cite{Weisse/2006,Ferreira/2015} as shown below. We note that the equivalence between Landauer-type and Kubo approaches to linear-response transport is well established, and we refer the interested reader to Refs. \cite{Fisher_Lee_81,Baranger/1989,Nikolic/2001} for additional details. To make use of the spectral machinery, we start by expanding the TBGF in terms of   Chebyshev polynomials of the first kind \cite{Boyd/2000}. To this end, we apply the  linear transformation $\hat{h}=(\widehat{H}-E_{+}\mathbf{1})/E_{-}$, with $E_{\pm}=(E_{>}\pm E_{<})/2$,  $\mathbf{1}$ is the identity operator defined on the Hilbert space of the lattice and $E_{>(<)}$ indicates the largest (smallest) eigenvalue of $\widehat{H}$. Note that this procedure maps the eigenvalues of the Hamiltonian onto the canonical interval of the Chebyshev polynomials i.e., $\mathcal{I}=[-1,1]$. Likewise, the Fermi energy variable is transformed according to $\varepsilon\equiv(E-E_{+})/E_{-}$.  To estimate the end points, $E_{\pm}$, we use a power method \cite{vonMises/1929}, and a `safety factor' is included to ensure that no spectral weight falls outside $\mathcal{I}$. This is achieved by means of a simple uniform re-scaling, $E_\pm \rightarrow (1+\alpha_{\text{SF}}) E_\pm$, with $\alpha_{\text{SF}}>0$ (in this work we use $\alpha_{\text{SF}}=0.1$).

In terms of the rescaled quantities introduced above, the imaginary part of the full TBGF admits the following Chebyshev decomposition \cite{Weisse/2006} 
\begin{equation}
 \textrm{Im }\widehat{\mathcal G}(\varepsilon)= \frac{2}{\pi\sqrt{1- \varepsilon^2}}\sum_{m=0}^{\infty} \frac{T_m(\varepsilon)}{(\delta_{m,0}+1)}\widehat {T}_{m}(\hat h),
\label{eq:KPM}
\end{equation}
where $T_{m}(\varepsilon)$ are Chebyshev polynomials of the first kind, and the  operators $\widehat T_{m}(\hat{h})$ ($m \in \mathbb Z^+$) satisfy the Chebyshev recurrence relations: $T_{0}(\hat{h})=\mathbf{1}$, $T_{1}(\hat{h})=\hat{h}$, and
\begin{equation}
\widehat T_{m+1}(\hat{h})=2\hat{h}\widehat T_{m}(\hat{h})-\widehat T_{m-1}(\hat{h}).
\label{eq:Chebyshev_recurrence}
 \end{equation}
 By virtue of these relations, Eq.~(\ref{eq:KPM}) and hence Eq.~(\ref{eq:KuboEq}) can be computed by means of an efficient iterative scheme based on computations of Chebyshev moments (see Sec. \ref{sec:Conductance} for details). Once the Chebyshev expansion [Eq.~(\ref{eq:KPM})] has been evaluated to the desired precision, the TBGF of the original system is obtained by a simple rescaling 
 $\textrm{Im } \widehat{\mathcal{G}}(E)=E_{-}^{-1} $\textrm{Im }$ \widehat{\mathcal G} (\varepsilon)$. 
\subsection{\label{sec:ChebyPoly}CAP and modified Chebyshev polynomials}
Next, we discuss the handling of the finite-size contacts in our implementation. As customary, the leads should be sufficiently large to behave as proper reservoirs of electrons. In practice, this is a demanding computational task, especially in nanostructures with large unit cells, such as the case of the TBG system of interest to this work. In order to reduce the computational overhead, we make use of a CAP approach \cite{Muga/2004, Manolopoulos/2004}. The CAP is a phenomenological damping term, 
\begin{equation}
\widehat{\Sigma} \equiv -i \widehat{\Gamma} = \textrm{diag}\,\, \{-\widehat{i\Gamma}_{L},0,-i\widehat{\Gamma}_{R}\}
\end{equation}
included in the Green's function that generates absorption of propagating waves across the contacts, thus minimizing reflections and emulating the behavior of a semi-infinite contact. The explicit form of the CAP self-energy for the setup in Fig. \ref{fig:schemeTBG} can be obtained by means of the Wentzel–Kramers–Brillouin semiclassical approximation as detailed in Ref. \cite{Manolopoulos/2004}, and is discussed  in Sec. \ref{sec:CAP}. Here, it is important to recognize that the presence of a self-energy term in the TBGF invalidates the  Chebyshev expansion (\ref{eq:KPM}) (this is the very reason why a standard self-energy formulation describing semi-infinite leads is avoided in our spectral approach), but can be conveniently handled by means of \textit{modified Chebyshev polynomials}, $\widehat{Q}_{n}(\hat{h},\hat \gamma)$ ($n \in \mathbb Z^+$), which are functions of the rescaled Hamiltonian, $\hat h$, and the damping operator, $\hat \gamma$. This  technique, originally devised for scattering calculations in molecular systems \cite{Mandelshtam/1995/1, Mandelshtam/1995/2}, allows to reconstruct the CAP Green's function, $\widehat{\mathcal G}^{\textrm{CAP}}\equiv(E-\widehat{H}-\widehat{\Sigma})^{-1}$, via the modified recurrence relations: $\widehat{Q}_0(\hat{h})  = \mathbf{1}$, $\widehat{Q}_1(\hat{h})   = e^{-\hat{\gamma}}  \hat{h}$ and
    \begin{align}
    \widehat{Q}_{m+1}(\hat{h})  = 2e^{-\hat{\gamma}}\hat{h} \widehat{Q}_{m}(\hat{h}) -e^{-2\hat{\gamma}}\widehat{Q}_{m-1}{(\hat{h})}
    \label{eq:modifiedQ}
   \end{align}
\noindent The formal relation between $\hat{\gamma}$---the main feature of the new recursion rule that fully encapsulates the effects of the CAP---and the original damping operator, $\widehat \Gamma $, is derived in the Appendix A for clarity.   
\subsection{\label{sec:Conductance}CAP-Chebyshev conductance algorithm}
To evaluate Eq. (\ref{eq:KuboEq}), the TBGF of the device is approximated by means of the modified Chebyshev polynomials introduced above. To reduce the cost associated with the trace operation in Eq. (\ref{eq:KuboEq}), we make use of a stochastic trace evaluation technique \cite{Iitaka_04}. It consists of replacing the exact trace by the average  expectation value over an ensemble of random vectors $\Ket{z_r}$ as follows
\begin{equation}
 G(\varepsilon)=\frac{4e^2}{h L_S^2 } \frac{1}{R}\sum_{r=0}^{R-1} \Bra{z_r}\hbar \hat{v}_x\hspace{1mm} \textrm{Im}\,\widehat{\mathcal G}(E)  \hspace{1mm}  \hbar \hat{v}_x  \hspace{1mm} \textrm{Im}\, \widehat{\mathcal G} (E)\Ket{z_r},
\label{eq:stochastic_trace}
\end{equation}
\noindent where $|z_r\rangle=\sum_{i=1}^{N} \chi_r |i\rangle$ is a vector with random amplitude on each lattice site, $N$ is the total number of orbitals and $ \chi_r $ are random (real) variables satisfying white-noise statistics (i.e., $\overline{ \chi_r}=0$ and  $\overline{\chi_r \chi_r^\prime}=\delta_{r,r^\prime}$, where the bar denotes the average over the random vector ensemble and $\delta_{r,r^\prime}$ is the  Kronecker delta symbol). The relative error in this approach scales favourably as $1/\sqrt{NR}$ \cite{Weisse/2006}, insofar as the operator being traced remains sparse \cite{Ferreira/2015}. As a rule of thumb, we set the number of random vectors such that $R\times N$ is on the order of $10^8$, which will afford us high accuracy in the evaluation of the conductance.

Next, we expand the TBGFs in Eq. (\ref{eq:stochastic_trace}) in terms of the modified Chebyshev polynomials [Eq. (\ref{eq:modifiedExpansion})] to obtain the $M$-order spectral approximation to $G(E)$. Following Ref. \cite{Ferreira/2015}, it is convenient to express the conductance as follows  
\begin{equation}
 G^{\text{CAP}}_M(\varepsilon)= \frac{4e^2}{h L_S^2 } \sum_{r=0}^{R-1}  \langle \phi_+^{(r)} (\varepsilon)| \phi_-^{(r)}(\varepsilon) \rangle, 
 \label{eq:Conductance_expansion}
 \end{equation}
with the single-shot vectors
\begin{align}
 &|\phi_+^{(r)}(\varepsilon)\rangle = \sum_{m=0}^{M-1}f_m(\varepsilon) \widehat{Q}_m(\hat{h})\hat{v}_x\Ket{z_r}, \label{eq:Evecs1}\\
 &|\phi_-^{(r)}(\varepsilon)\rangle = \sum_{m=0}^{M-1}f_m(\varepsilon)\hat{v}_x\widehat{Q}_m(\hat{h})\Ket{z_r},
 \label{eq:Evecs2}
 \end{align}
 \noindent where
 \begin{align}
 f_m(\varepsilon)=k_m\frac{(2-\delta_{m,0})T_m(\varepsilon)}{\sqrt{1-\varepsilon^2}}
 \label{eq:auxfnm}
\end{align} 
\noindent and {$k_m$} are Jackson kernel coefficients \cite{Weisse/2006} introduced to suppress Gibbs oscillations generated by the truncation of the formal infinite  series in Eq. (\ref{eq:KPM}). In this work, we will use $M$ up to $20000$ which corresponds to a smearing of the delta functions (i.e. energy resolution) of $\delta E = \pi E_{-}/M \approx 1 \textrm{ meV}$ at the band center. 

The single shot vectors are constructed on the fly via a sequence of standard matrix-vector multiplications. First, by defining the vector $\ket{z^{m}_r}\equiv\widehat{Q}_m(\hat{h})\Ket{z_r}$, Eq. (\ref{eq:modifiedQ}) can be used to yield the sequence
    \begin{align}
    \ket{z^{m}_r}=2e^{-\hat{\gamma}}\hat{h}\ket{z^{m-1}_{r}}-e^{-2\hat{\gamma}}\ket{z^{m-2}_{r}} 
    \label{eq:singleshot1}
   \end{align}
\noindent which is initiated with $\ket{z^{0}_r}\equiv\ket{z_r}$  and $\ket{z^{1}_r}=e^{-\hat{\gamma}}\hat{h}\ket{z_r}$. This process is iterated to obtain the $M$-th order approximation defined as $|\phi_-^{(r)}(\varepsilon)\rangle=\sum_{m=0}^{M-1}f_m(\varepsilon)\hat{v}_x\ket{z^{m}_{r}}$. A similar procedure, but with starting vectors $\ket{\overline{z}^{0}_r}=\hat{v}_x\ket{z_r}$ and $\ket{\overline{z}^{1}_r}=e^{-\hat{\gamma}}\hat{h}\hat{v}_x\ket{z_r}$, yields the remaining single shot vector $|\phi_+^{(r)}(\varepsilon)\rangle=\sum_{m=0}^{M-1}f_m(\varepsilon)\ket{\overline{z}^{m}_{r}}$. 
   
The numerical determination of the single-shot vectors, $| \phi_{\pm}^{(r)}(\varepsilon) \rangle$, is the most demanding part of the CAP-Chebyshev algorithm. However, the complexity of this approach grows only linearly (see Fig. \ref{varCAP}(b)) with the system size because the relevant matrices in the Chebyshev iteration,  $\hat h$ and $\gamma$, are sparse. All together, the number of operations required by the algorithm scales as $N\times E \times R \times M$, where $E$ is the number of energy points being considered.


The single-shot algorithm adapted here to the Landauer problem provides a particularly efficient scheme for evaluation of Fermi surface terms in linear response theory as it scales linearly with the number of Chebyshev iterations. In contrast, the standard spectral evaluation of  Eq. (\ref{eq:Conductance_expansion}) would require  the evaluation of a total of $M^2$ expansion moments of the type $\mu_{nm}=\text{Tr} [\widehat{T}_n(\hat h) \hat v_x \widehat {T}_m(\hat h) \hat v_x]$ (or $\tilde \mu_{nm}=\text{Tr} [\widehat{Q}_n(\hat h) \hat v_x \widehat{Q}_m(\hat h) \hat v_x]$, with the modified Chebyshev polynomials). Instead, the vectors Eqs. (\ref{eq:Evecs1})-(\ref{eq:Evecs2}) are constructed in parallel by means of a matrix-vector multiplication scheme exploiting the recursive rule in Eq. (\ref{eq:modifiedQ}) as well as the sparseness of the Hamiltonian matrices. The total number iterations in this approach is $2M$ (as opposed to $M^2$ in a full-spectral calculation based on the explicit evaluation of Chebyshev moments). For more details on the single-shot approach and its efficient numerical implementation for large systems, we refer the reader to the supplementary material of Ref. \cite{Ferreira/2015}. In this work, we compute the conductance for 100 single-shot energy points in parallel. The computational time for a single random vector realization of the largest system simulated (i.e., $L=300$ nm and $W=100$ nm with $M=20000$ Chebyshev iterations) is approximately 27 minutes using a  NVIDIA Tesla K80 graphics card. The memory cost is low (approximately 10 GB).  

\subsection{\label{sec:CAP} CAP implementation and benchmark}
 Here, we use a CAP adapted from Refs. \cite{Manolopoulos/2004, Manolopoulos/2002},  previously used in quantum transport simulations of graphene devices  \cite{Gaetano/2018,Andelkovic_18, Munoz/2012}. It has the following form

\begin{align}
&\Gamma(x)=- \frac{4E_{\textrm{min}}}{c^2}y(x_s')  \hspace{4mm}\textrm{where:}\label{finalCAP}  \\ 
&y(x'_s)=\left[\frac{1}{(1-x'_s)^2}+\frac{1}{(1+x'_s)^2}-2 \right],  \\
& x'_{s}=(x_{s}-x_{0})/L_C . 
\end{align}

\noindent Here, $x'_{s}$ is the $s$-th site's relative position inside the contact, which starts at $x_{s_0}=x_{0}$ and has a length $L_C$. In practice, $E_{\textrm{min}}$ becomes an adjustable parameter that will determine the lowest electron energy that the CAP can absorb and $c=2.66206$ is a numerical constant. Once a contact length is defined, $L_C$, an optimal value of $E_{\textrm{min}}$ can be found by numerical experimentation. Far from such optimal value, the simulations will show spurious oscillations (finite size effects), in addition to oscillations due to wave reflections at the contact terminations.

\begin{figure}%
\begin{center}
\resizebox{8.6cm}{!} {\includegraphics{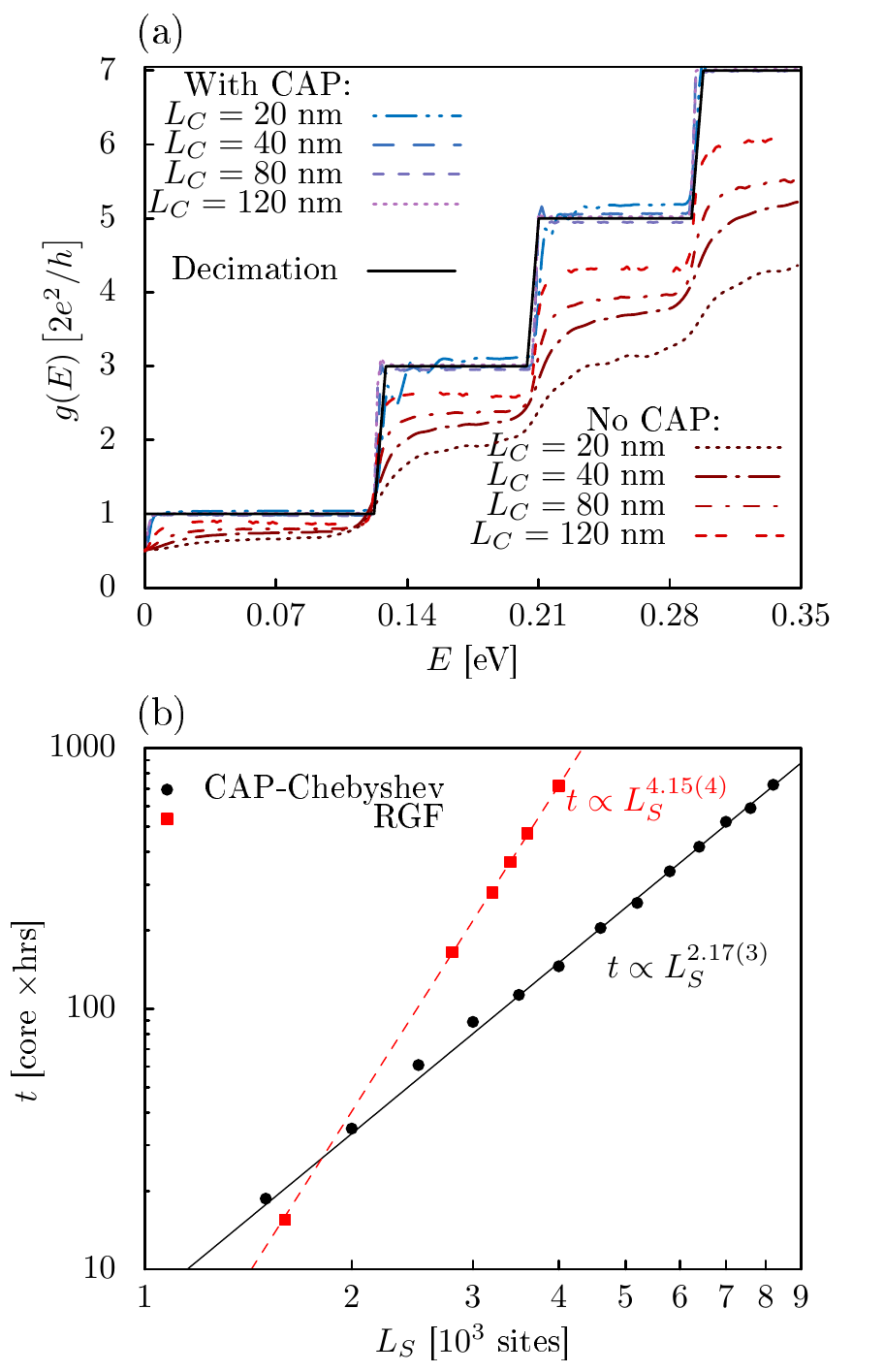}}
 \caption{(a) Two-terminal conductance as a function of energy for selected values of $L_C$ with and without a  CAP. The result obtained with the standard RGF approach \cite{Lewenkopf_13} is shown (black solid line) for comparison. We use $W=$ 24 nm and $L_S=20$ nm. The number of Chebyshev iterations increases gradually with the contact size from $M=10000$ to $ M=16000$ as the contact size increases, while the number of random vectors is kept fixed with $R=400$. (b) Scaling of the total computational cost (core $\times$ hours) as a function of device length $L_S$ for a single-shot evaluation of the conductance using the Chebyshev (circles) and RGF (squares) methods.}
 \label{varCAP}%
\end{center}
\end{figure}

Once the functional form of $\widehat{\Gamma}$ is determined using the  above recipe, its dimensionless version $\hat \gamma$ appearing in the rescaled Green's function is obtained via Eq. (\ref{eq:gamma_explicit}). Since we are interested in the low-energy properties of TBG, in what follows we use the approximation $\hat \gamma \simeq \textrm{asinh}(i \widehat \Gamma/E_-)$ (see Appendix A), which simplifies the Chebyshev recursion somewhat by rendering  $\hat \gamma$ energy-independent. 

We validate our CAP approach on a zigzag graphene nanoribbon (see Fig. \ref{varCAP}). A standard tight-binding model with a nearest-neighbour hopping energy of $t=-2.7$ eV was used for this purpose. In this system, the energy-dependence of the conductance exhibits the familiar sequence of ballistic conductance steps \cite{Lewenkopf_13}. The comparison against the RGF method (Fig. \ref{varCAP}(a)) shows that the CAP performance improves very quickly as the contact size increases  \cite{Manolopoulos/2004}. The spectral results already approximate reasonably well those  obtained by the standard RGF technique for $L_C = 20$ nm. The accuracy improves gradually with contact size, with the results for  $L_C \ge 120$ nm being virtually indistinguishable from the exact conductance profile. Note that in the absence of a CAP, the impact of finite-size effects severally affect the quality of the simulation (this can be seen by the significant reduction of the conductance below its quantized values in Fig. \ref{varCAP}(a)). These results  highlight the advantage of the CAP approach when handling finite contacts. 

Next, we carried out systematic calculations to compare the computational effort of the CAP-Chebyshev approach with that of the standard RGF technique. To this end, we used nanoribbon geometries with  $W=L_S$, such that the total number of orbitals scales proportionally to $L_S^2$.  A decimation method is used to evaluate the Green's functions of the leads within the RGF approach, while the  CAP-Chebyshev method employs finite contacts with the same size as the central region. The CPU-time scaling with the respect to the system size is shown in Fig. \ref{varCAP} (b). The spectral method is undoubtedly efficient for studies of large systems, with a computational effort roughly linear with the total number of orbitals in the system (i.e., $t_{\text{CPU}}\propto N \propto L_S^2$). On the other hand, the RGF approach exhibits an approximate quadratic dependence, which is very demanding in general for  large systems. It is important to note that the number of Chebyshev iterations in the spectral approach must increase with the system size so that the energy levels are fully resolved \cite{Ferreira/2015,KITE_2020}. Here, we used $M= \text{int} (2.4 N_x)$, where $N_x$ is the number of sites along the length $L_S$. This results in $M=20000$ (or, equivalently, an energy resolution on the order of $1$ meV) for the largest system in our benchmark ($N_x=8\times 10^3)$, and yields accurate results for the conductance.

The observed CPU-time scaling in the spectral method is completely consistent with the algorithm complexity outlined in Sec. \ref{sec:Methodology} B. In fact, the number of floating-point operations (per energy point) grows as $\mathcal O \propto N\times M \times R $, and so should the CPU-time too.  Given that $M\propto N^{1/2}$ (in order to achieve a suitable, size-dependent energy resolution), we expect $t_{\text{CPU}} \propto R \times N^{1.5}$. Here, the stochastic trace evaluation of the conductance [Eq.~(\ref{eq:stochastic_trace})] plays a crucial role. Because its relative error  scales roughly as $1/\sqrt{N R}$ (the exact scaling is also sensitive to $M$ \cite{Ferreira/2015,KITE_2020}), the required number of random vectors decreases with the system size (until it saturates at $R=1$ for very large $N$). For our system, $R= \text{int} (a/N_x)$ with $a=O(10^5)$ yields convergent results with the same precision across all system sizes. This in turn implies $\mathcal O \propto N$, and hence the observed CPU-time scaling ($t_{\text{CPU}} \propto N$). We note in passing that for simulations of very large systems (e.g., relevant to capture the diffusive regime of graphene flakes of realistic size containing billions of orbitals \cite{Ferreira/2015}), $R=1$ suffices to achieve high accuracy, and thus $t_{\text{CPU}} \propto N^{1.5}$ would be expected in that limit. Further details about our benchmark are provided in the Appendix B.

\section{\label{sec:TBG} TBG ballistic conductance}
\subsection{\label{sec:pristine}The pristine case}
Having outlined the real-space Chebyshev approach to quantum transport and demonstrated its performance in a simple graphene device, we now apply it to the study of TBG. The twisting of van der Waals heterostructures has recently provided a novel tuning mechanism in condensed matter physics \cite{TBLG_tw_Carr2017,TBLG_tw_Cao2018,TBLG_tw_Isobe2018,TBLG_tw_Ribeiro2018,TBLG_tw_Andrei_20}. The interference pattern arising from two off-kilter graphene sheets creates a moiré supercell, which can be viewed as a new overarching crystal cell of the material. Because the moiré period can be much larger than the original lattice scale of graphene (especially for small twist angles), the study of TBG devices is computationally challenging.  Furthermore, an accurate tight binding description of TBG has to handle a multitude of nearest-neighbours hoppings \cite{TBLG_TB_Moon_12}, which adds complexity to the numerical calculations. Therefore, the evaluation of the conductance in realistic large-unit-cell TBG systems has been out of reach of standard approaches in quantum transport, including the popular recursive Green's function method \cite{Bahamon/2019}.

From the perspective of quantum transport, there are a few works on TBG \cite{Andelkovic_18,Bahamon/2019,Pelc,Sanz_20}. These works focused on the twist-angle dependence of the minimal CNP conductivity  \cite{Andelkovic_18} and transmission properties of small junctions \cite{Bahamon/2019,Pelc,olyaei2020ballistic,Sanz_20}. The comparisons in Fig. \ref{varCAP}(a)  indicate that the CAP-Chebyshev approach is well suited to tackle large systems, thus overcoming the limitations of standard approaches to two-terminal conductance calculations. The device geometry employed in this work is depicted in Fig. \ref{fig:schemeTBG}. It comprises two  armchair graphene nanoribbons twisted by an angle $\theta$ and separated by an interlayer distance $d_0=0.335$ nm. Each nanoribbon has its own left and right contact regions, with a length of $L_C$ and the same width $W$ as the central system. In the calculations reported below, the length of the contacts is equal to the length of the central region, and thus the total linear size of the device is $L=3L_C$.  

To model the electronic properties of the TBG system, we employ a single-orbital tight-binding model with hopping terms parameterized as follows \cite{TBLG_TB_Moon_12, Koshino/2012/2, SlaterKoster/1954}:
\begin{equation}
  -t_{i,j}=V_{pp\pi}\left[ 1-\left( \frac{\bm{d}_{ij}\cdot\bm{e}_z}{d_{ij}} \right)^2\right] + V_{pp\sigma}\left( \frac{\bm{d}_{ij}\cdot\bm{e}_z}{d_{ij}}  \right)^2,
 \label{eq:hoppings} 
\end{equation}
\noindent where $V_{pp\pi(\sigma)}=V_{pp\pi(\sigma)}^0 \text{ exp}\left( -\frac{d_{ij}-a_0(d_0)}{\delta} \right)$,  $\bm{d}_{ij}=\bm{r}_i -\bm{r}_j$ is the vector connecting two sites (here, $d_{ij}=|\bm{d}_{ij}|$), and $a_0=0.142$ nm and $d_0=0.335$ nm are carbon–carbon distance in graphene and interlayer distance, respectively. Moreover, $V_{pp\pi}^0=-2.7$ eV and $V_{pp\sigma}^0=0.48$ eV are the bilayer graphene nearest-neighbor intralayer and interlayer hopping integrals, respectively, $\delta=0.32 a_0$ is the hopping decay length and $\bm{e}_z$ is the unit vector normal to the TBG plane. In order to reproduce the main features of the single-particle electronic structure, it is imperative to go beyond the nearest neighbours approximation \cite{Koshino/2012/2, Bahamon/2019}. Thus, in this work we incorporate all interlayer and intralayer neighbours  within a sphere of radius $4a_0$ centered at $\bm{r}_i$. It is important to mention that our TBG graphene Hamiltonian describes both the device region and the finite contact regions. This confers a practical advantage for achieving a plausible description of electronic transport in TBG devices because incoming wavefunctions in our approach already possess twist properties, which allows us to overcome momentum mismatch between incoming and outgoing states \cite{Bell_2014}. 

\begin{figure}[ht!!]%
\begin{center}
\resizebox{8.6cm}{!}{\includegraphics{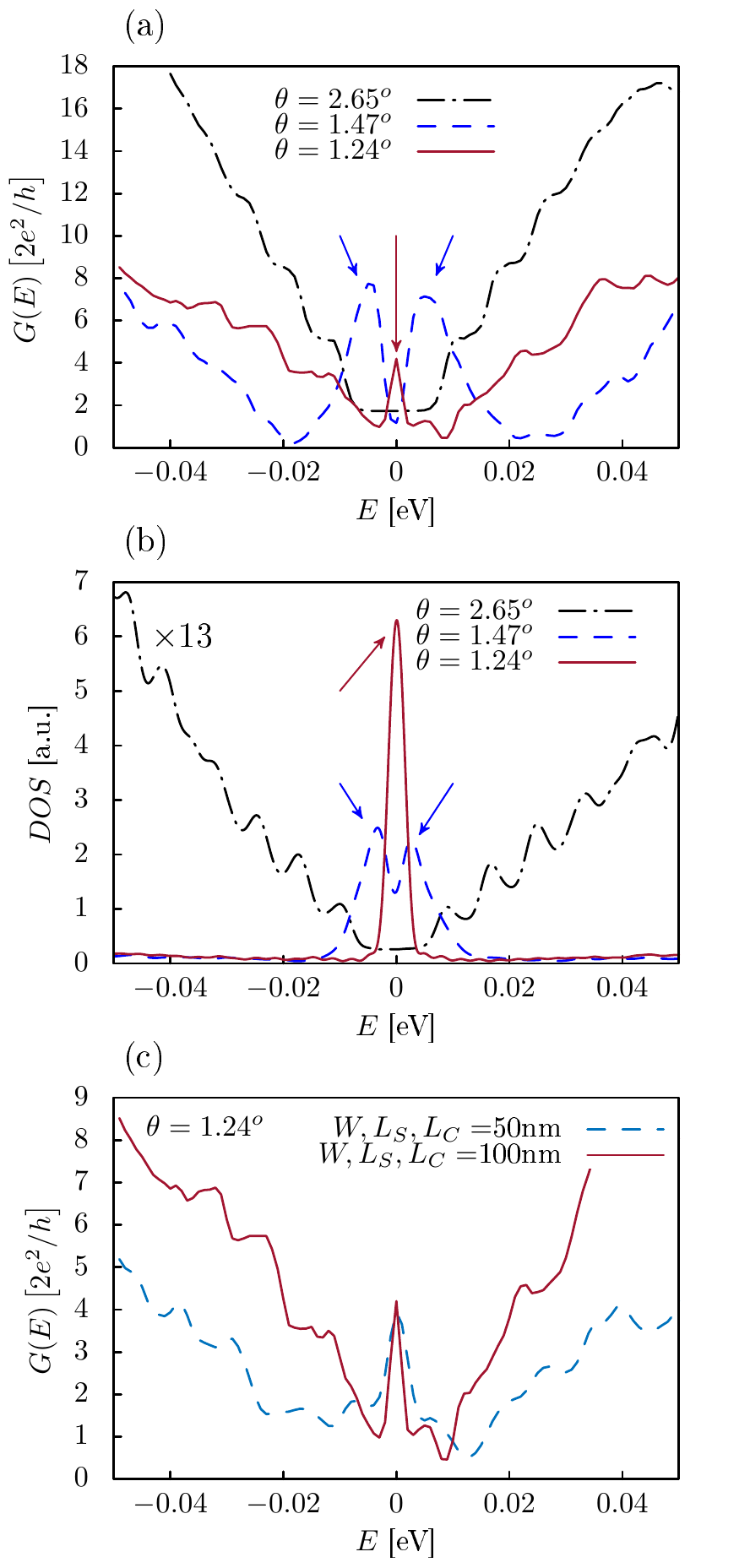}}
\caption{Electronic properties of armchair TBG devices for selected twist angles. (a) Fermi energy dependence of the zero-temperature conductance. Parameters: $M=20000$, $R=150$, and $L_C=L_S=W=100$ nm. (b) Density of states for the systems considered in the top panel calculated with the kernel polynomial method \cite{Ferreira/2011} (same parameters as above). (c) Conductance profile for two samples with different dimensions as indicated in the legend and $\theta=1.24^\circ$. Parameters: $M=14000$ and $R=250$ for the $W=50$ nm case and $M=20000$ and $R=150$ for the $W=100$ nm case.}
\label{fig:TBGcond}
\end{center}
\end{figure}

We investigated devices composed of TBG at selected commensurate twist angles. Let us first summarize the key results obtained with the CAP-Chebyshev approach. For large twist angles, the layers are effectively decoupled and the conductance is approximately twice that of a single-layer nanoribbon (this is discussed in Sec. \ref{sec:Disorder}, alongside with the effect of twist-angle disorder). For intermediate twist angles, the conductance plateaus are still clearly visible [see Fig. \ref{fig:TBGcond}(a) for a device with $\theta=2.65^{\circ}$], but have a decreased width (compared to the single nanoribbon) due to moiré pattern effects. As the twist angle is reduced below $2^{\circ}$, one begins to probe the effects of strong interlayer coupling. In this regime, the ballistic conductance plateaus are smoothed out due to strong channel mixing caused by elastic scattering between the layers. As a result, the conductance away from the charge neutrality point (CNP) attains much lower values than those of the untwisted armchair ribbon counterpart. More interestingly, the quantum transport near the CNP becomes dominated by quasi-localized states in the AA-stacked regions of the bilayer. These states produce van Hove singularities (VHSs) at the vicinity of the CNP. A spectral study of the density of states is reported in Fig. \ref{fig:TBGcond}(b), where the emergence of peaks in the density of states at small twisting can be seen for twist angles of $1.24^{\circ}$ and  $1.47^{\circ}$. Likewise, the conductance in this regime displays prominent peaks at the VHS locations indicated by arrows in Fig. \ref{fig:TBGcond}(a). For the device with $\theta=1.47^{\circ}$, a group of low-dispersion electronic states with a bandwidth of only $\sim 40$ meV is expected to appear near the CNP based on electronic calculations for the bulk system \cite{TBLG_TB_Moon_12}. This is consistent with the electronic properties exhibited by our nanostructures. Note that the width of the central double peak structure in Fig.  \ref{fig:TBGcond}(b) is around $40$ meV.

As the lowest twist angle studied here ($\theta=1.24^\circ$), the above features merge into a single peak at the CNP (indicated by the red arrow), with the conductance acquiring a stable value close to $8 e^2/h$. This isolated conductance peak is a result of the residual dispersion from the moir\'e minibands. Note that a truly flat energy band would yield zero conductance in a non-interacting picture. The observed CNP conductance seems to be robust with respect to variations in the device dimensions. Indeed, Fig. \ref{fig:TBGcond}(c) shows that the conductance for smaller devices show the same peak height, even though these devices have a smaller number of transverse modes and, consequently, display lower values of the conductance away from the CNP. We replicated the behavior of the smaller device simulated using a standard Landauer method with a wide band model for the contacts \cite{Bahamon/2019,PhysRevB.88.235433}. 
\subsection{\label{sec:Disorder}Twist Angle disorder}
Lastly, we take advantage of our real-space approach to examine the role played by twist angle disorder. There are few theoretical works that study this crucial kind of disorder in TBG \cite{Wilson/2020,PhysRevResearch.2.033458,PhysRevB.104.075144} and even fewer addressing the quantum transport problem \cite{PhysRevB.104.075144}. In Ref. \onlinecite{PhysRevResearch.2.033458} an effective two band model was used to calculate the transmission across twist angle domains, while in Ref. \onlinecite{PhysRevB.104.075144} a 1D analog (double-wall carbon nanotube with angle disorder) of TBG is studied. To incorporate angle disorder in our mesoscopic TBG systems, we make use of a simple toy model \cite{Wilson/2020}. The top layer of the device region is divided in four sections, each having their own rotation angle sample from a box distribution $[\bar\theta-\Delta \theta, \bar\theta+\Delta \theta ]$, where $\theta$ is the  average twist angle and $\Delta \theta$ is the maximum deviation from the central value. For each region the  interlayer hoppings are recalculated with the displaced positions of the top layer sites according to Eq. (\ref{eq:hoppings}), while the intralayer hoppings are kept at their unperturbed values.  
\begin{figure}[ht!!]%
\begin{center}
\resizebox{8.6cm}{!} {\includegraphics{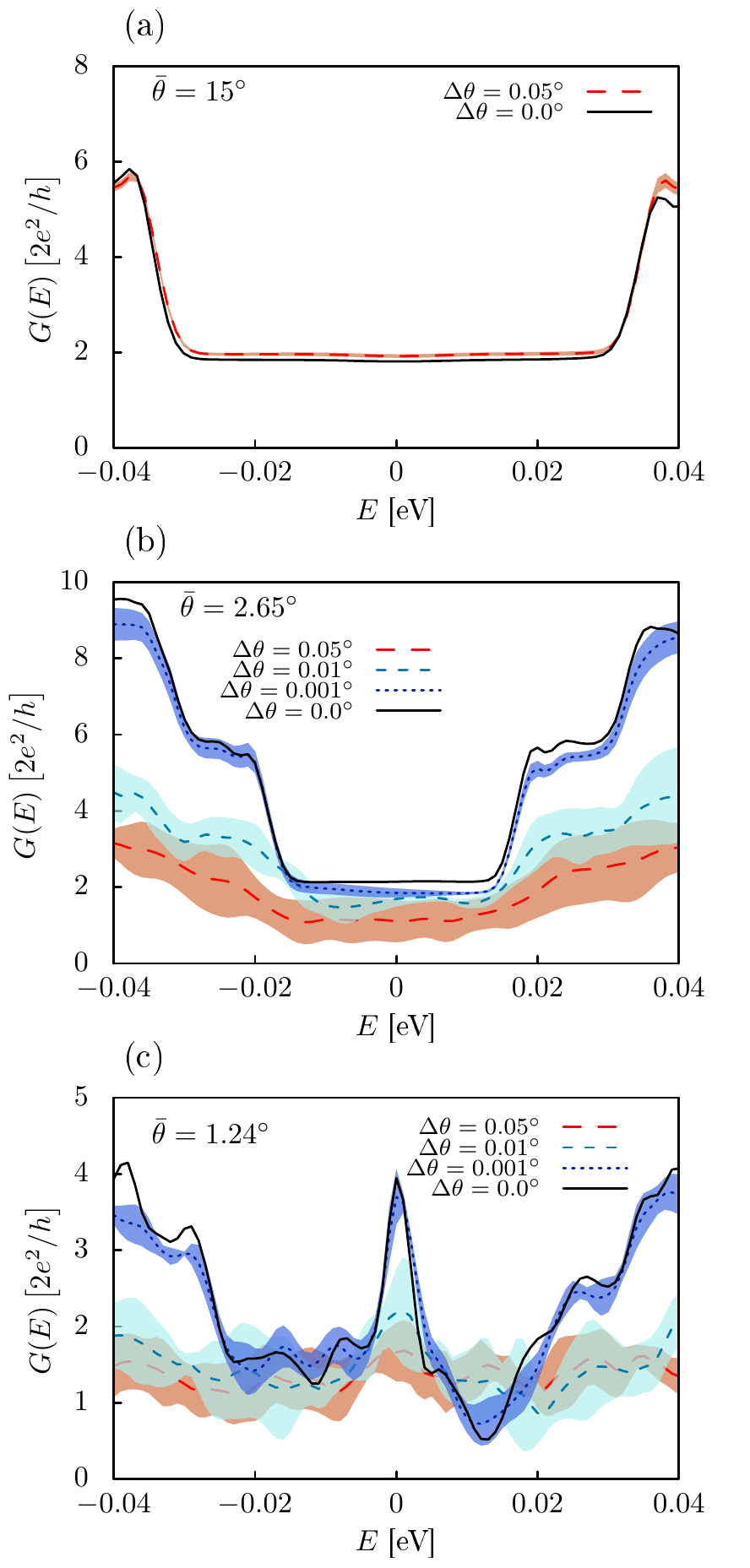}}
\caption{Conductance curves in the presence of twist angle disorder for systems with an average twist angle  $\bar \theta=1.24^\circ$ (a), $\bar \theta=2.65^\circ$ (b), and $\bar \theta=15^\circ$ (c), with the shaded bands showing the statistical uncertainty. Parameters: $L=L_C=L_S=50$ nm, $M=14000$ and $R=250$. }
\label{fig:TBGdisorder}
\end{center}
\end{figure}
Figure \ref{fig:TBGdisorder} shows the averaged conductance   for small ($\bar\theta=1.24^\circ$), intermediate ($\bar\theta=2.65^\circ$) and large ($\bar\theta=15^{\circ}$) average twist angles. Three disorder strengths are investigated, corresponding to $\Delta \theta=0.001^\circ,~0.01^\circ~\text{and}~0.05^\circ$. The conductance is averaged over five twist-angle disorder realizations and is represented by a continuous line, while the standard deviation is shown as a shaded band. For comparison, we have shown the clean case ($\Delta \theta=0.0^\circ$). As expected, for weak angle disorder ($\Delta \theta=0.001^\circ$) the main characteristics of the conductance curves are preserved, which include the peak at the CNP for the samples with $\bar \theta=1.24^\circ$  and the well-defined conductance steps for  $\bar \theta=2.65^\circ$. It is also evident that the conductance only shows small deviations from the clean case, especially near the CNP. On the other hand, for mild disorder ($\Delta \theta=0.01^\circ,0.05^\circ$), the conductance shows large variation between different samples (except for the trivial case with $\theta=15^{\circ}$ given that the layers are effectively decoupled). The observed effects on the averaged conductance are consistent with the broadening and extinction of the van Hove singularities, as observed in  Ref. \cite{Wilson/2020}. In the range of twist angles probed here, the average conductance is seen to decrease with angle disorder very quickly, which is perhaps surprising considering that the largest $\Delta \theta$ considered is only $0.05^\circ$. This behavior can be rationalized in terms of the strong dependence of the spectral properties upon the twist angle. In fact, even minute variations in twist angle are sufficient to suppress the van Hove peaks of the local density of states \cite{Wilson/2020}.

These results show that twist angle variations as small as $0.4\%$ can have a massive impact on the quantum transport properties of mesoscopic TBG devices. This is in stark contrast to the behavior of the conductance in TBG nanotubes \cite{PhysRevB.104.075144}, whose quantum transmission properties exhibit  strong resilience to angle disorder. Our disorder framework is however too simplistic to capture quantitatively the impact of realistic twist-angle disorder landscapes in a standard TBG nanostructure (i.e., it is composed only of  four squared twist angle domains with sharp boundaries). We note that these limitations, while severe, can be overcome by improving our microscopic model. We leave a more in-depth study of the impact of twist-angle disorder on the conductance of realistic systems for future work.

\section{\label{sec:Conclusions}Concluding Remarks} 
In this work, we developed and validated a real-space CAP-Chebyshev approach to two-probe conductance calculations in the linear transport regime. As an application, we studied mesoscopic TBG devices with a focus on small twist-angle systems and the transport signatures of low-dispersion energy bands in the vicinity of the CNP. This is one of the first applications of the spectral method to the Landauer transmission problem \cite{Joao/2020,Yu_2020} and the first, to our knowledge, to formulate a direct expansion of the two-terminal conductance in terms of Chebyshev polynomials. The use of complex absorbing potentials, and associated modified Chebyshev polynomials, has allowed us to alleviate the computational cost of simulating (large) leads that behave as proper electron reservoirs. This makes the hybrid CAP-Chebyshev approach particularly efficient for conductance calculations of large disordered two-dimensional systems.   
Our study of TBG nanostructures of mesoscopic lateral dimensions (length up to $300$ nm and width up to $50$ nm) has shown that the ballistic conductance depends strongly upon the twist angle as well as the degree of angle disorder in the sample. Broadly speaking, three transport regimes were identified. At large twist angles ($\theta \gg  1^\circ$),  the transport channels of each layer are effectively decoupled and the conductance is twice that of a monolayer device (this is consistent with the bulk dc transport characteristics of TBG \cite{Andelkovic_18}). For intermediate angles ($10^\circ \gtrsim \theta \gtrsim 2^\circ$), the conductance curves exhibit well-defined quantization steps with a step width modulated by moiré pattern effects. For $\theta \lesssim 2^\circ$, the conductance steps are washed out due to a strong channel mixing caused by coherent interlayer scattering. Finally, for twist angles approaching the largest magic angle ($\theta_m \simeq 1.1^\circ$), two conductance peaks located approximately symmetrically on either side of the CNP. This feature is traced back to a pile-up of energy states near the CNP, which in bulk samples leads to the well-known van Hove singularities in the density of states. These peaks merge into a single peak at the CNP when the twist angle approaches $1.24^\circ$, the smallest twist angle in our study. We have also briefly addressed the issue of twist disorder, a type of spatial inhomogeneity that is ubiquitous in realistic systems \cite{Uri_20}. According to our preliminary calculations using a simple model of twist-angle disorder, the quantum transport characteristics of TBG devices are surprisingly sensitive to abrupt changes of twist angle across domains. In strucutres with a small twisting ($ \bar \theta \lesssim 2^\circ$), variations of only $0.01^\circ$ suffice to reduce the conductance by a factor of two.   

While preparing this manuscript, we became aware of a related numerical study by Ciepielewski and co-workers \cite{Ciepielewski:2022}, where transport signatures of van Hove singularities in mesoscopic TBG devices with sizes comparable to ours are reported.
\section{Acknowledgements}
A.F. acknowledges support from a Royal Society University Research Fellowship. SGdC and DAB acknowledge support form the Brazilian Nanocarbon Institute of Science and Technology (INCT/Nanocarbon), CAPES-PROSUC (grant no. 88887.510399/2020-00, Doctorate degree), FAPESP (grant 18/07276-5), CAPES-PRINT (grant no. 88887.575078/2020-00, Sandwich doctorate), CAPES-PRINT (grant no. 88887.310281/2018-00),  CNpQ (309835/2021-6) and Mackpesquisa. The supercomputer time was provided by the high-performance computing cluster of the Mackenzie Presbyterian University (https://mackcloud.mackenzie.br). We acknowledge F. M. O. Brito for proofreading the final version of the manuscript. SGdC also acknowledges the hospitality of the School of Physics, Engineering and Technology at the University of York, U.K., where this work was completed.
\section{Appendix A\label{sec:Appendix-A}}
 The modified Chebyshev recursion relation is obtained from the following identity \cite{Mandelshtam/1995/2}
    \begin{align}
     &\frac{1}{\textrm{cos}(\textrm{acos}(\varepsilon)-i\hat{\gamma})-\hat{h}} \nonumber \\ 
     &= \frac{1}{\textrm{sin}(\textrm{acos}(\varepsilon)-i\hat{\gamma})}\sum_{n=0}^{\infty}  c_n e^{-in\hspace{1mm}\textrm{acos}(\varepsilon)}\widehat{Q}_{n}(\hat{h}),
     \label{eq:modifiedExpansion}
   \end{align}
with $c_n=2-\delta_{n,0}$. Equation (\ref{eq:modifiedExpansion}) provides a polynomial expansion for the Green's function of the system with the CAP:
 \begin{align}
 \widehat{\mathcal G}&  \equiv \frac{1}{E_-}
 \frac{1}{\textrm{cos}(\textrm{acos}(\varepsilon)-i\hat{\gamma})-\hat{h}}
\label{eq:equiv-1} \\  
 & =\frac{1}{E-(\widehat{H}+i\widehat{\Gamma})}.\label{eq:equiv-2}
\end{align}
The expressions (\ref{eq:equiv-1})-(\ref{eq:equiv-2}) yield a relationship between the dimensionless damping operator, $\hat{\gamma}$,  appearing in the rescaled Green's function and the absorbing potential in the original Green's function, $\widehat{\Gamma}$, i.e.
   \begin{align}
     \widehat{i\Gamma}=&E_{-}[ \textrm{cos}(\textrm{acos}(\varepsilon))(1-\textrm{cosh}(\hat{\gamma})) \nonumber \\
      &-i\hspace{1mm}\textrm{sin}(\textrm{acos}(\varepsilon))\textrm{sinh}(\hat{\gamma})]. \label{eq:gammaDefine}
   \end{align}  
For the class of Hermitian damping operators we focus on (resulting in a purely imaginary CAP self-energy), the above relation can be easily inverted to yield 
\begin{equation}
\hat{\gamma}=\textrm{asinh}\left(-\frac{\widehat{\Gamma}}{i\hspace{1mm}E_{-}\hspace{1mm}\sqrt{1-\varepsilon^2}}\right).  
\label{eq:gamma_explicit}
\end{equation}
At low energies, $|\varepsilon|\ll1$, the above expression can be safely approximated as $\hat \gamma \simeq \textrm{asinh} (i \widehat \Gamma / E_-) $. 

\section{Appendix B\label{sec:Appendix-B}}
The simulations in our benchmark are run on CPU nodes equipped with Intel Xeon 6138 20-core 2.0 GHz processors. The number of random vectors ($R)$ in the CAP-Chebyshev method is determined by a simple convergence analysis. The convergence parameter is defined as  $\delta_R=|(G^{(R)}(\varepsilon)-G^{(R-1)}(\varepsilon))/G^{(R-1)}(\varepsilon)|$, where $G^{(R)}$ is the single-shot conductance obtained with $R$ random vectors; see Eq.~(\ref{eq:stochastic_trace}). A result is considered converged when this metric falls below $0.03$. For the  range of system sizes in our simulations, this is satisfied by selecting $R = \text{int} (1.64 \times 10^5 / N_x)$.  Combined with the dependence of number of Chebyshev moments with the system size (see main text), this results in $t_{\text{CPU}}\propto N_x^{2.2}$. In contrast, the RGF method shows a scaling proportional to $N_x^{4.2}$, in accord with previous studies (e.g., using the KWANT numerical package \cite{Groth_2014}). For the smaller systems simulated ($N_x < 1.5\times 10^3$ sites)  the situation is inverted, with the RGF technique exhibiting better performance (see Fig. \ref{varCAP} (b)).
\bibliography{BibMaterials} 
\end{document}